\documentclass[10pt,pra,twocolumn,floatfix]{revtex4}
\usepackage{amssymb}
\usepackage{amsfonts}
\usepackage{amssymb}
\usepackage{amsmath}
\usepackage[dvips]{graphicx}
\usepackage{color}
\usepackage{appendix}

\usepackage{graphicx}
\usepackage{dcolumn}
\usepackage{bm}

\usepackage{graphicx}

\usepackage{hyperref}

\usepackage{mathrsfs}
\usepackage{isomath}
\usepackage{amsthm}
\usepackage{epstopdf}
\usepackage{txfonts}

\begin{document}

\title{Remote preparation of optical cat states based on Gaussian entanglement}

\author{Dongmei Han$^{1}$, Fengxiao Sun$^{2}$, Na Wang$^{1}$, Yu Xiang$^{2,3}$, Meihong Wang$^{1,3}$, Mingsheng Tian$^{2}$, Qiongyi He$^{1,2,3,*}$, and Xiaolong Su$^{1,2}$}

\email{qiongyihe@pku.edu.cn}
\email{suxl@sxu.edu.cn}

\affiliation{$^{1}$State Key Laboratory of Quantum Optics and Quantum Optics Devices, 
Institute of Opto-Electronics, Shanxi University, Taiyuan, 030006, China\\ 
$^{2}$State Key Laboratory for Mesoscopic Physics, School of Physics, Frontiers Science Center for Nano-optoelectronics, Peking University, Beijing, 100871, China\\
$^{3}$Collaborative Innovation Center of Extreme Optics, Shanxi University,
Taiyuan, Shanxi, 030006, China\\
$^{4}$Peking University Yangtze Delta Institute of Optoelectronics, Nantong, Jiangsu, 226010, China
}

\begin{abstract}
Remote state preparation enables one to prepare and manipulate quantum state non-locally. As an essential quantum resource, optical cat state is usually prepared locally by subtracting photons from a squeezed vacuum state. For remote quantum information processing, it is essential to prepare and manipulate optical cat states remotely based on Gaussian entanglement, which remains a challenge. Here, we present experimental preparation of optical cat states based on a remotely distributed two-mode Gaussian entangled state in a lossy channel. By performing photon subtraction and homodyne projective measurement at Alice's station, an optical cat state is prepared remotely at Bob's station. Furthermore, the prepared cat state is rotated by changing Alice's measurement basis of homodyne detection, which demonstrates the remote manipulation of it. By distributing two modes of the two-mode Gaussian entangled state in lossy channels, we demonstrate that the remotely prepared cat state can tolerate much more loss in Alice’s channel than that in Bob’s channel. We also show that cat states with amplitudes larger than $2$ can be prepared by increasing the squeezing level and subtracting photon numbers. Our results make a crucial step toward remote hybrid quantum information processing involving discrete- and continuous-variable techniques.
\end{abstract}

\maketitle

\section{Introduction}

With the development of quantum communication and quantum network, it becomes possible for a user without the ability of preparing quantum state to obtain quantum resources. Generally, there are several options to achieve this goal, such as direct state transmission, remote state preparation (RSP)~\cite{Lo2000,Paris2003continuous}, and quantum teleportation~\cite{Noriyuki2011}, respectively. RSP enables one to create and control a quantum state remotely based on shared entanglement. Compared with the direct state transmission, where a prepared quantum state is directly transmitted to the user through a lossy channel, RSP offers remote control of quantum state and intrinsic security~\cite{Pogorzalek2019,leung2003}. Compared with quantum teleportation, RSP does not need joint measurement, requires less classical communication~\cite{Aru2000}, and offers the ability to manipulate quantum state remotely. 

Schr\"{o}dinger cat states play important roles in both fundamental physics and quantum information, such as exploring the boundary between quantum and classical physics \cite{Schrodinger1935,HarocheRMP2013,Arndt2014}, quantum computation \cite{JeongPRA2002,RalphPRA2003,LundPRL2008,Tipsmark2011}, quantum communication \cite{Ulanov2017,Sychev2018,vanEnkPRA2001,jonas2013} and quantum metrology \cite{Kira2011,JooPRL2011,Gilchrist2004}. Free-propagating optical cat states have attracted much attention attributed to their weak interaction with the environment, which is beneficial to quantum information processing. Up to now, most of the reported optical cat states are prepared locally by subtracting photons from squeezed vacuum states~\cite{Ourjoumtsev2006,Neergaard2006,kentaro2007,Takahash2008,Thomas2010,Sychev2017,zhang2021}, which combines the discrete-variable technology and continuous-variable resource~\cite{Ulrik2015}. However, this method sets a barrier for the users since it requires the ability of creating squeezed states and performing a non-Gaussian operation. 

Recently, RSP has been applied to prepare non-Gaussian states ~\cite{walschaers2020remote,xiang2022,shuheng2022}, which demonstrate the connection between remotely prepared Wigner negativity and quantum steering. As for a special kind of non-Gaussian states, cat states have also been prepared by RSP based on a two-photon Fock state \cite{Ourjoumtsev2007}, a two-photon N00N state~\cite{Ulanov2016}, hybrid discrete- and continuous-variable entanglement~\cite{Laurat2018,Hacker2019} and optomagnetic entanglement~\cite{Fengxiao2021}. In most previous experiments~\cite{Ulanov2016,Laurat2018}, the RSP of cat states demonstrates the generation of a non-Gaussian state from a non-Gaussian entangled resource. Compared with preparing the non-Gaussian entangled resource, such as the N00N state and hybrid entangled state, Gaussian entangled states can be prepared deterministically and present scalability~\cite{Christian2012,Yokoyama2013,Roslund2014,Chen2014,Mikkel2019,Asavanant2019}. However, it still remains a challenge to experimentally prepare and manipulate optical cat states remotely based on Gaussian entanglement. 

Here, we experimentally demonstrate the preparation of cat states at a distant node based on a distributed Gaussian entangled state in a lossy channel. Alice, who has the ability to perform photon subtraction, and Bob, who doesn't, share a two-mode squeezed state (TMSS) remotely. By implementing photon subtraction and homodyne projective measurement on Alice's state, Bob's state collapses to a cat state conditionally. An optical odd cat state with amplitude of $\sim 0.65$ and fidelity of $\sim 0.67$ is created at a generation rate of $1$ kHz by projecting on phase quadrature at Alice's station when the transmission efficiency on Bob's mode is $0.9$. Then, the cat state is rotated for 90 degrees by converting the projective measurement to amplitude quadrature at Alice's station, which demonstrates remote manipulation of the prepared state. Moreover, remote preparation of optical cat states is achieved when Alice's or Bob's state is transmitted through a lossy channel. We also show that the amplitudes of the prepared cat states can be increased by subtracting more photons from a TMSS with optimum squeezing. In principle, this scheme can remotely prepare odd or even cat states by subtracting odd or even photon numbers at Alice's station. Thus, our result provides a new method to remotely generate and manipulate optical cat states.

\begin{figure}[tb]
\centering
\includegraphics[width=\linewidth]{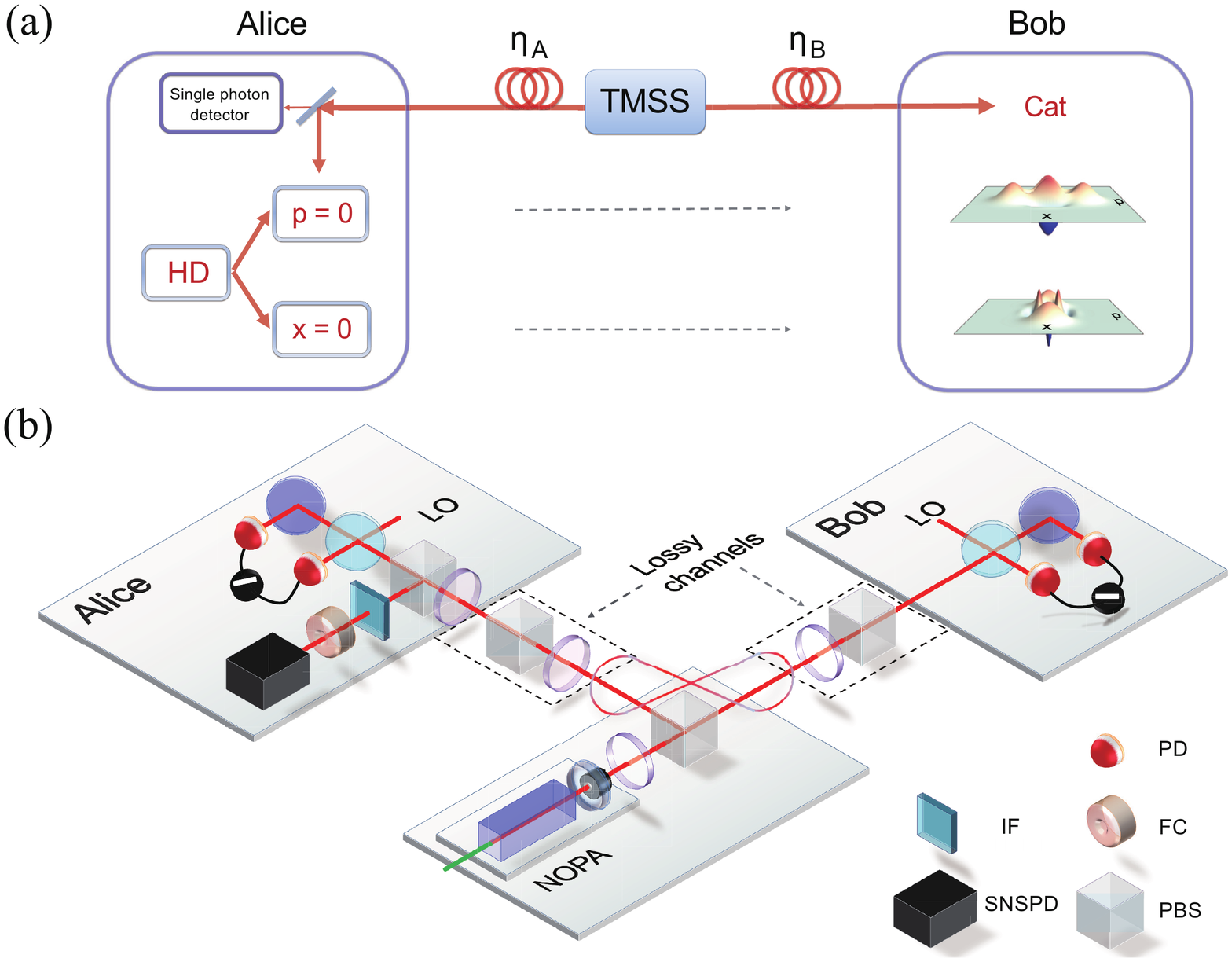}
\caption{Schematic and experimental setup. a) The principle of the experiment. Alice implements photon subtraction by using a single-photon detector and homodyne projective measurement on one mode of the TMSS. A cat state or rotated cat state for $90$ degrees is created at Bob's station conditioned on the measurement results of $p_{A}=0$ or $x_{A}=0$. b) Experimental setup. 
Bob's lossy channel is simulated by the combination of a half-wave plate (HWP) and a polarization beam splitter (PBS). TMSS, Two-mode squeezed state; HD, Homodyne detector; NOPA, Non-degenerate optical parametric amplifier; PD, Photodiode; IF, Interference filter; FC, Filter cavity; SNSPD, Superconducting nanowire single photon detector; LO, Local oscillator.}
\label{fig1}
\end{figure}

\section{The Principle}

As shown in Figure~\ref{fig1}a, a TMSS of the form $|\psi_0\rangle_{AB}=\frac{1}{\cosh r}\sum^{\infty}_{m=0}\tanh^{m}r|m,m\rangle_{AB}$ is prepared, where $r$ is the squeezing parameter, and then two modes of the entangled state are sent to Alice and Bob through lossy quantum channels with transmission efficiencies of $\eta_{A}$ and $\eta_{B}$ respectively. Alice performs photon subtraction on her state and measures the photon-subtracted state with a homodyne detector (HD). By measuring the quadrature $\hat{x}_A^\theta=(\hat{a}_A e^{-i\theta}+\hat{a}_A^{\dagger}e^{i\theta})/\sqrt{2}$ and projecting the output to $x_A^\theta=0$, where $\hat{a}_{A}$ is the annihilation operator and $\theta$ is a general phase, an odd cat state is remotely prepared at Bob's station. Especially, $\hat{x}_A^\theta$ corresponds to the amplitude quadrature ($\hat{x}_A$) for $\theta=0$ and the phase quadrature ($\hat{p}_A$) for $\theta=\pi/2$. 
And the eigenstate $|x_{A}^{\theta}\rangle_{A}$ with corresponding eigenvalue $x_A^{\theta}$ is expressed by the inner product with the photon number state $|m\rangle$
\begin{equation}
	{_{A}}\langle x_{A}^{\theta}|m \rangle_{A}=\frac{e^{-im\theta}}{\sqrt{2^{m}m!\sqrt{\pi}}}e^{-(x_A^\theta)^2/2}H_m(x_A^\theta),
	\label{eigenstate}
\end{equation}
where $H_m(x)$ is the Hermite polynomial. 

For an arbitrary quadrature measurement $\hat{x}_A^\theta$ with the outcome chosen as $x_A^\theta=0$, the ideal conditional state obtained at Bob's station becomes $|\varphi \rangle_{B}={}_{A}\langle 0^{\theta}|\hat{a}_A|\psi_0\rangle_{AB}\propto\sum_{m=1}^{\infty} \sqrt{{m}/{2^{m-1}(m-1)!}}e^{-i(m-1)\theta}\tanh^m r$
$H_{m-1}(0)|m\rangle_{B}$. Since $H_{2k}(0)=(-2)^{k}(2k-1)!!$ and $H_{2k+1}(0)=0$, the state can be simplified as
\begin{equation}
|\varphi \rangle_{B} \propto|r,\theta\rangle_B-|-r,\theta\rangle_B,
\end{equation} 
with $|r,\theta\rangle_B\propto \sum_{m=0}^{\infty}\frac{m!!}{\sqrt{m!}}e^{-i m(\theta-\pi/2)}{\rm{tanh}}^{m}r |m\rangle_B$ (see Appendix A for more details). Hence, if Alice measures the phase quadrature ($\hat{p}_A$), Bob's state is similar to an odd cat state $|cat_{-}\rangle=(\vert\alpha\rangle-\vert-\alpha\rangle)/\sqrt{2(1-e^{-2\vert\alpha\vert^2}})$ with the real number $\alpha$. Ideally, the fidelity $F=\langle cat_{-}|\rho_{B}|cat_{-}\rangle$ between Bob's state $\rho_{B}=|\varphi \rangle_{BB}\langle\varphi|$ and the odd cat state $|cat_{-}\rangle$ reaches $\sim99\%$ with $3$ dB squeezing of the TMSS (see Appendix A for more details). If Alice measures the amplitude quadrature ($\hat{x}_A$), Bob's conditional state is similar to the state $(|i\alpha\rangle-|-i\alpha\rangle)/\sqrt{2(1-e^{-2\vert\alpha\vert^2}})$, which is equivalent to applying a rotation operation by 90 degrees $\hat{R}(\pi/2)$ on the odd cat state $|cat_{-}\rangle$. This indicates that Bob's cat state can be remotely manipulated by choosing the basis of Alice's homodyne projective measurement. 

\begin{figure*}[t]
\centering
\includegraphics[width=0.86\linewidth]{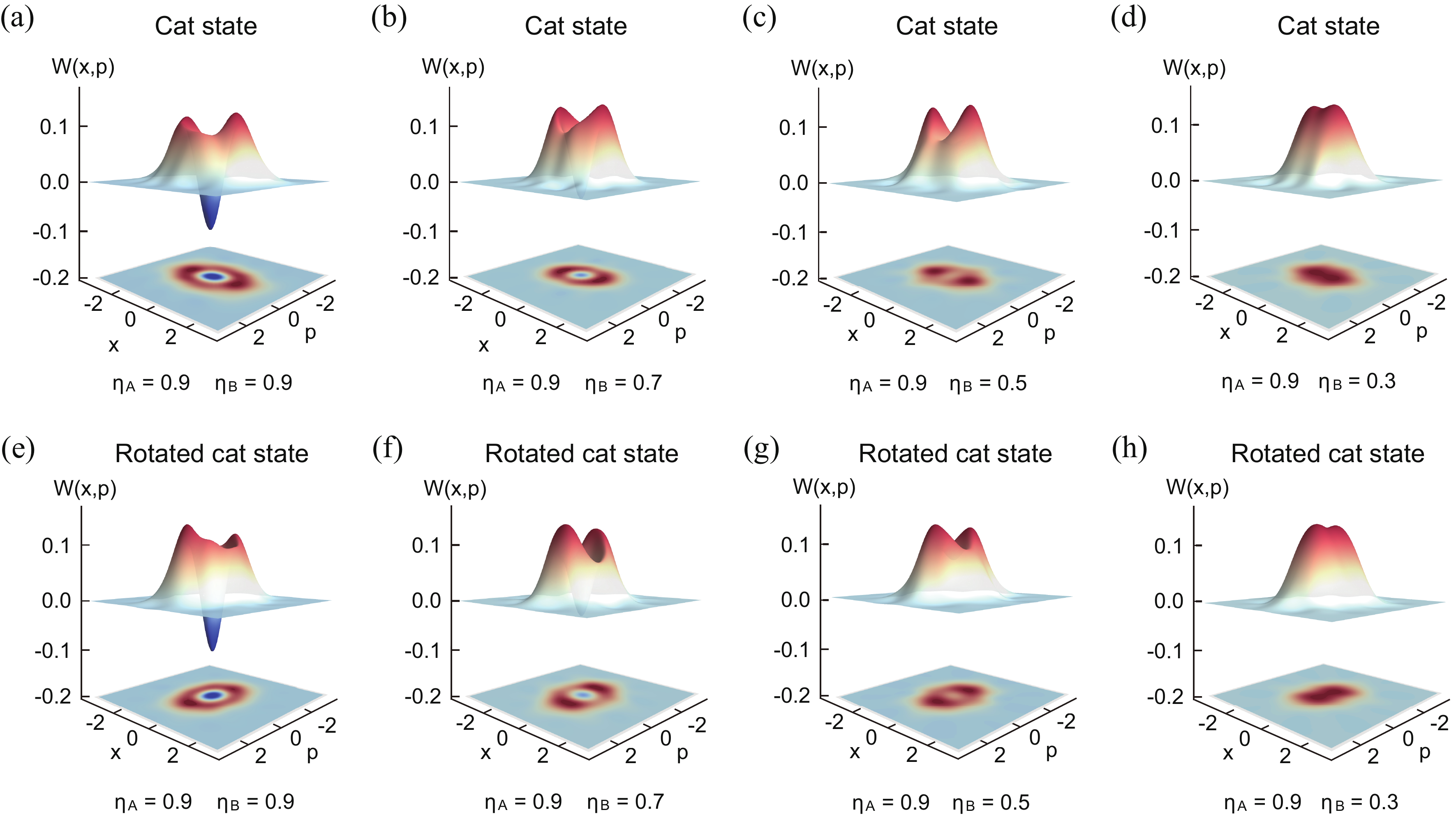}
\caption{Reconstructed Wigner functions and corresponding contour plots of prepared states at different transmission efficiency $\eta_{B}$. a), b), c), d) Cat states and e), f), g), h) rotated cat states for 90 degrees at different transmission efficiencies of Bob's mode. All results in the above plot are corrected with 90\% detection efficiency.}
\label{fig2}
\end{figure*}

\section{The Experiment}

As shown in Figure~\ref{fig1}b, when the non-degenerate optical parametric amplifier (NOPA) is working at amplification status, where the relative phase between seed beam (1080 nm) and pump beam (540 nm) is locked to 0, a TMSS with squeezing and antisqueezing levels of $-$3.2 dB and $+$4.2 dB is prepared when we inject 70 mW pump power into the NOPA.~\cite{liu2019,deng2021,wang2020} (see Appendix B for more details). To perform photon subtraction operation, Alice uses a variable beamsplitter composed of a half-wave plate (HWP) and a polarization beam splitter (PBS) to tap around $4\%$ of her mode towards the superconducting nanowire single-photon detector (SNSPD). An interference filter with 0.6 nm bandwidth together with a filter cavity with 400 MHz bandwidth are placed in front of the SNSPD to select the degenerate mode of the NOPA. To avoid the reflection from the front mirror of the filter cavity to the NOPA, an optical isolator is placed in front of the interference filter. An SNSPD with around $70\%$ detection efficiency, which only influences the generation rate of the prepared state, is used to detect the subtracted photons and the clicks of it are used to trigger the storage oscilloscope for the data recording of Alice's and Bob's homodyne detectors (HDs). Our experiment is conducted in the Locking-and-hold mode, where the seed beam is injected into the NOPA for the cavity locking during the locking period (around 50 ms), and it is chopped off to obtain the TMSS during the hold period (around 30 ms) (see Appendix B for more details). 

To realize the homodyne projective measurement, the amplitude quadrature $\hat{x}_{A}$ or phase quadrature $\hat{p}_{A}$ of the photon-subtracted state is measured by Alice's homodyne detector and then post selected with a selection width of  $\delta x < 0.05$ on $\hat{x}_{A}$ ($\hat{p}_{A}$). Alice and Bob record the output signals of their HDs simultaneously, and then Bob only keeps the corresponding data when Alice's quadrature values meet the selection condition. Bob performs quantum tomography to reconstruct the Wigner function $W(x,p)$ of his state.

\section{Results}

As shown in Figure~\ref{fig2}a and~\ref{fig2}e, optical cat states with two directions in phase space are prepared by projecting Alice's quadrature values at $p_A = 0$ and $x_A = 0$, respectively. In principle, cat states at Bob's station could be rotated in arbitrary directions with the form of $(|e^{i\theta}\alpha\rangle-|e^{-i\theta}\alpha\rangle)/\sqrt{2(1-e^{-2\vert\alpha\vert^2}})$ by changing the homodyne projection angle $\theta$ at Alice's station (see Appendix A for more details), which shows remote manipulation of directions of prepared cat states in phase space. Compared with the preparation of cat states in arbitrary directions based on photon subtraction from squeezed vacuum states, which requires squeezed vacuum states squeezed in arbitrary directions, the presented scheme relaxes this requirement on the quantum resource.

An optical cat state and a rotated cat state with amplitude $|\alpha|\sim0.65$ and the value of $W(0,0)\sim-0.10$ are obtained when transmission efficiency is $\eta_{A}=\eta_{B}=0.9$. The fidelity of the prepared optical cat state $F=\left\langle cat_{-}\right\vert {\rho}_{out}\left\vert cat_{-}\right\rangle $ is quantified by calculating the overlap between an ideal cat state $\left\vert cat_{-}\right\rangle$ and the experimentally reconstructed density matrix ${\rho}_{out}$. An optical cat state and a rotated cat state with the fidelity of $F\sim67\%$ are obtained, which is limited by the purity of the initial TMSS and loss, at the transmission efficiency of 0.9.

\begin{figure*}[tb]
\centering
\includegraphics[width=0.66\linewidth]{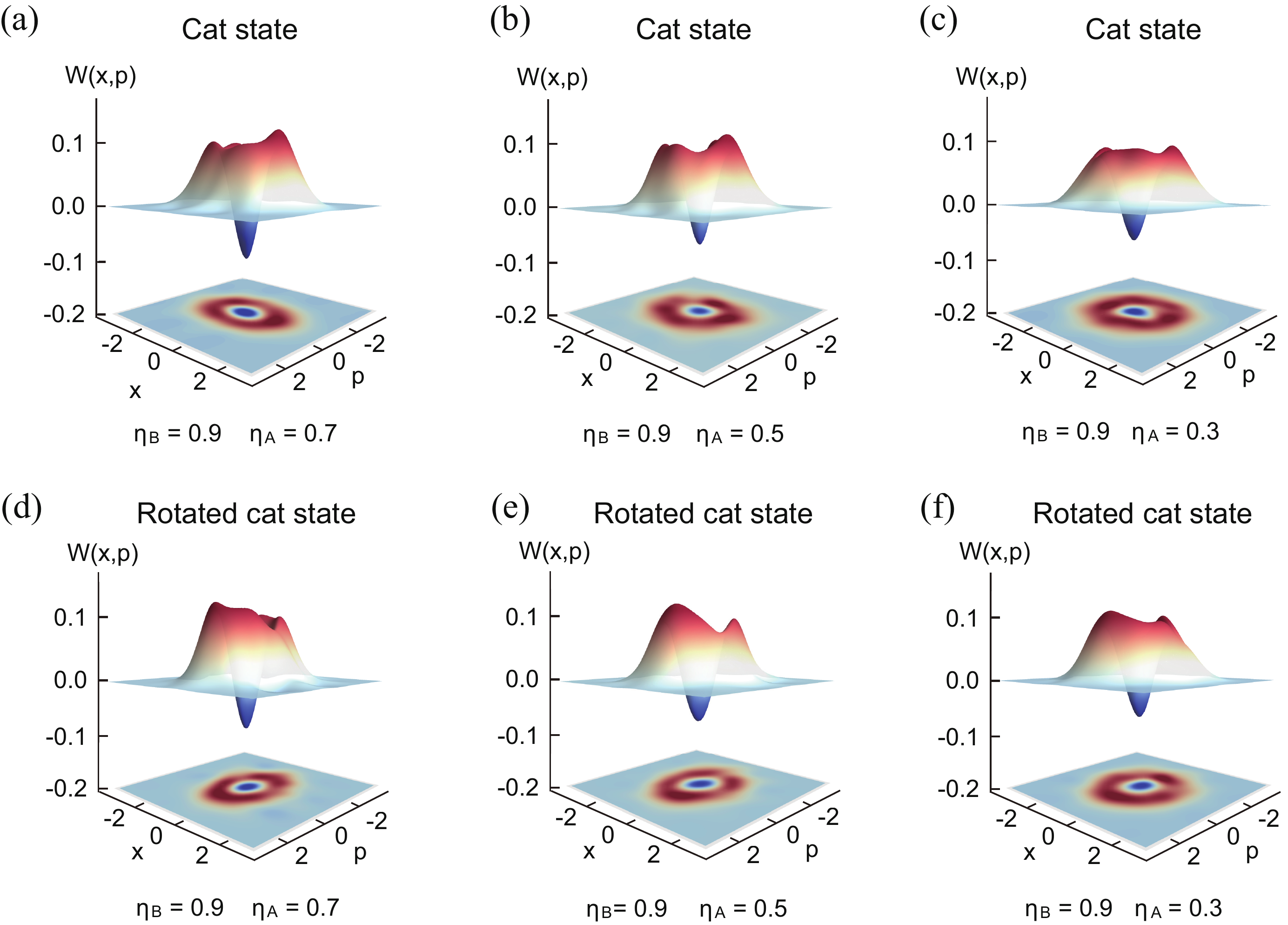}
\caption{Reconstructed Wigner functions and corresponding contour plots of prepared states at different transmission efficiency $\eta_{A}$.
a), b), c) Cat states and d), e), f) rotated cat states for 90 degrees at different transmission efficiencies of Alice's mode. All results in the above plot are corrected with 90\% detection efficiency.}
\label{fig3}
\end{figure*}

\begin{figure*}[t]
\centering
\includegraphics[width=0.8\linewidth]{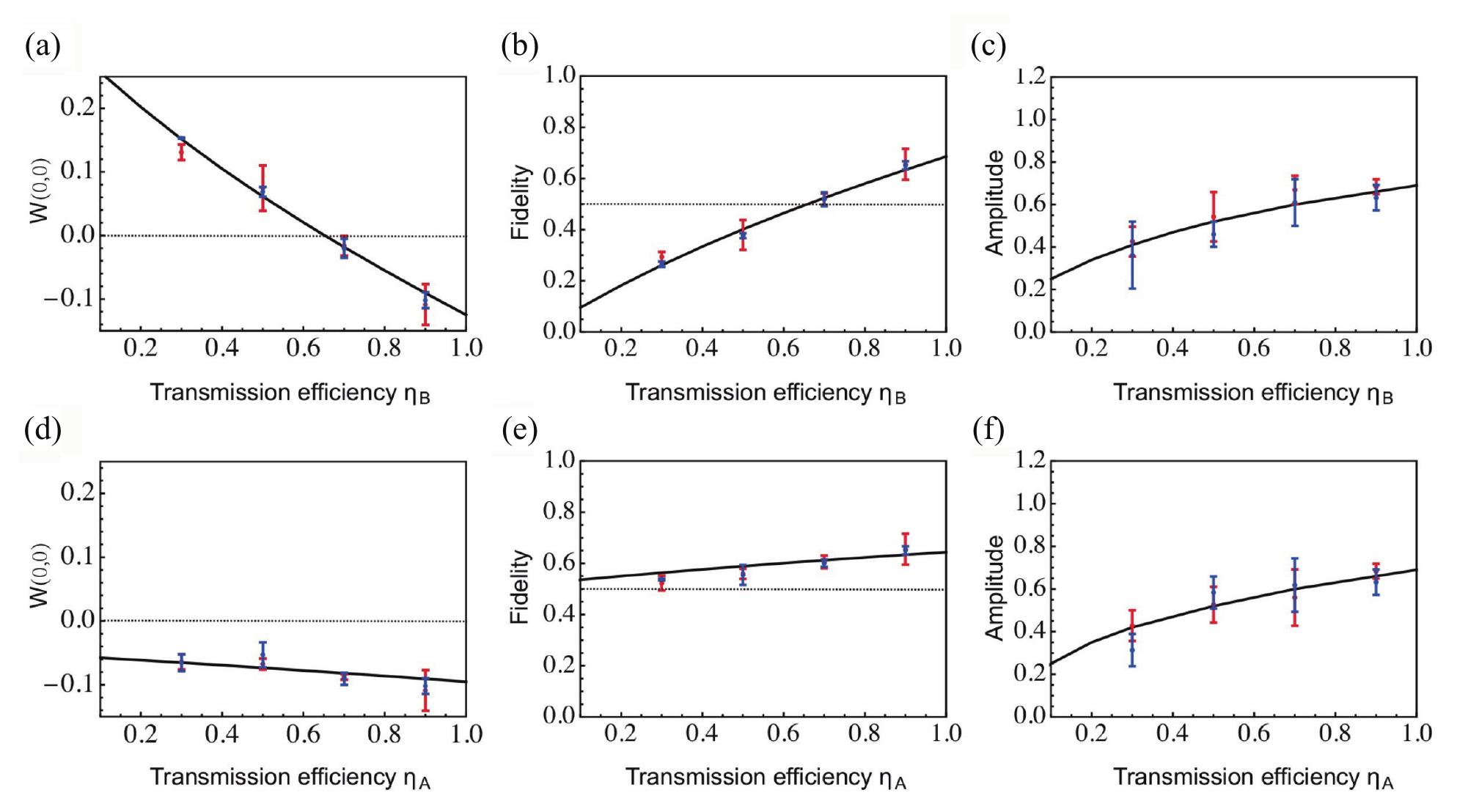}
\caption{Evolution of remotely prepared cat states in a lossy quantum channel. a) The Wigner negativity, b) fidelity, and c) amplitude of the remotely prepared cat states as a function of the transmission efficiency of Bob's mode. d) The Wigner negativity, e) fidelity, and f) amplitude of the remotely prepared cat states as a function of the transmission efficiency of Alice's mode. The red and blue data points represent cat states and rotated cat states for $90$ degrees, respectively. The error bars are obtained by the standard deviation of measurements repeated three times.}
\label{fig4}
\end{figure*}

\begin{figure*}[t]
\centering
\includegraphics[width=0.7\linewidth]{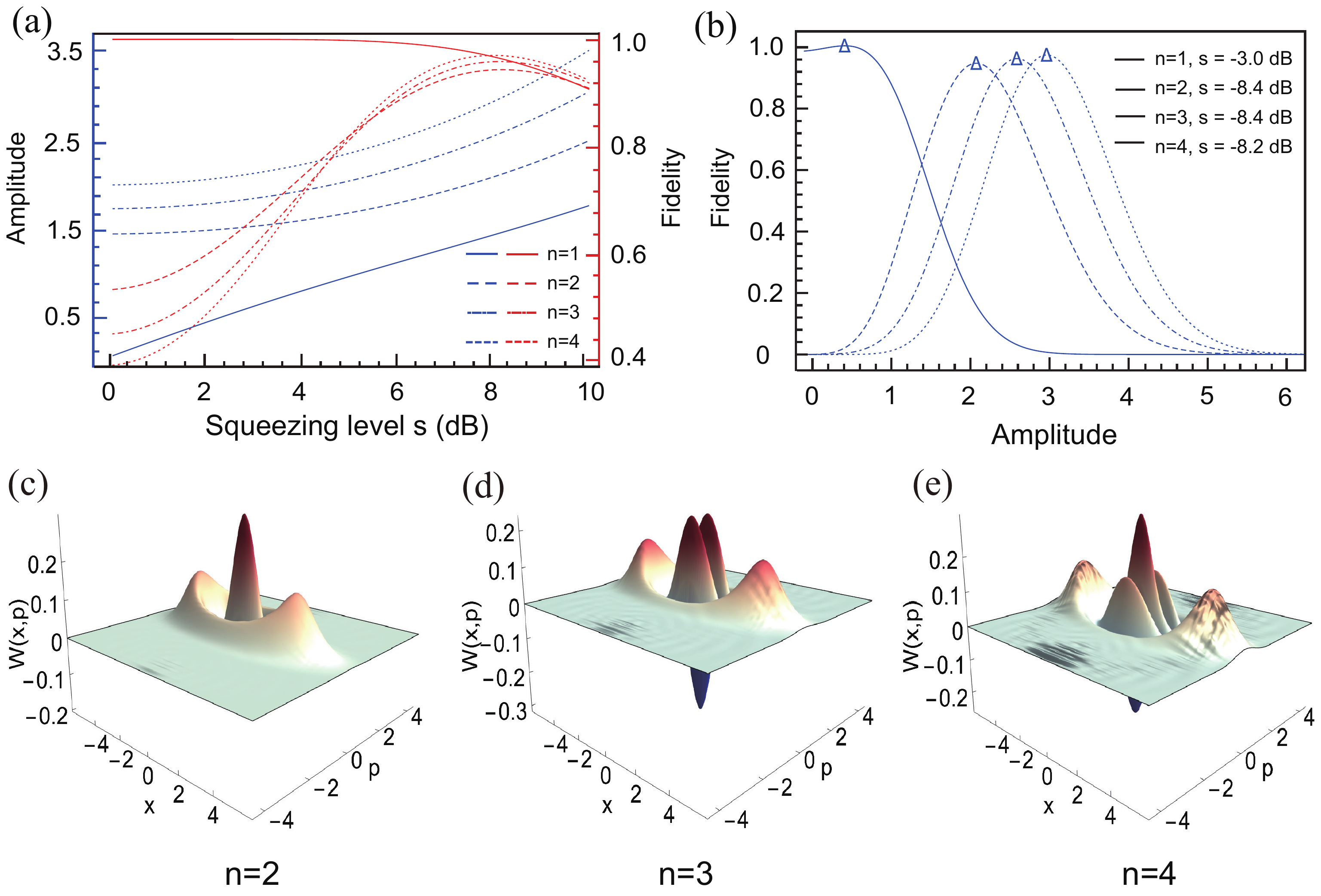}
\caption{Results for increasing the amplitude of the prepared cat states.
a) Dependence of amplitude and fidelity of Bob's cat state on the squeezing level of the TMSS for different subtracted photon numbers on Alice's mode. b) Dependence of fidelity on the amplitude of Bob's cat state with optimum squeezing levels for different subtracted photon numbers. The fidelity represents the overlap between Bob's state $\rho_{B}$ and the odd cat state $|cat_{-}\rangle$ for odd $n$, and the overlap between $\rho_{B}$ and the even cat state $|cat_{+}\rangle=(|\alpha\rangle+|-\alpha\rangle)/\sqrt{2(1+e^{-2|\alpha|^2})}$ for even $n$. The unit transmission efficiency is chosen in the calculation. c), d), e) Wigner functions of prepared cat states at Bob's station when two, three, and four photons are subtracted from Alice's mode with optimum squeezing levels of $-$8.4 dB, $-$8.4 dB and $-$8.2 dB, respectively.}
\label{fig5}
\end{figure*}

To present the tolerance of our scheme on channel loss, we simulate the transmission of Alice's and Bob's modes of the TMSS in lossy channels by changing the transmission efficiency in Alice's and Bob's quantum channels respectively. In the case of loss on Bob's mode, as shown in Figure~\ref{fig2}, the negative values of $W(0,0)$ of the prepared cat state and the rotated cat state are reduced when the transmission efficiency is varied from 0.9 to 0.3, which represents the decrease of the nonclassical feature. The negative part of the Wigner function vanishes at the transmission efficiency of 0.5 and 0.3.

In the case of loss on Alice's mode, as shown in Figure~\ref{fig3}, cat and rotated cat states are also obtained at Bob's station. The negative value of $W(0,0)$ of prepared states is reduced when the transmission efficiency is varied from 0.7 to 0.3, but it never disappears. Compared with the case of loss in Bob's mode (Figure 2),  the loss tolerance of our scheme in Alice's mode is better than that in Bob's mode. 

To quantify the characteristics of remotely prepared cat states in a lossy quantum channel, we present the value of $W(0,0)$, fidelities, and amplitudes of prepared states at different transmission efficiencies in Bob's and Alice's channels in Figure~\ref{fig4}. In the case of loss on Bob's channel [Figure~\ref{fig4}a, \ref{fig4}b, and \ref{fig4}c], it is obvious that the fidelity and amplitude of remotely prepared cat states are reduced with the decrease in transmission efficiency. When the transmission efficiency of Bob's mode surpasses $0.64$, the fidelity is larger than 50\%, and negativity of the Wigner function appears, which indicates that cat states are successfully prepared remotely within the transmission distance of around $9$ km (considering a loss rate of $0.2$ dB/km in the fiber channel). 

In the case of loss on Alice's channel [Figure~\ref{fig4}d, \ref{fig4}e, and \ref{fig4}f], it is obvious that the negativity of the Wigner function, fidelity, and amplitude of remotely prepared cat states is reduced with the decrease of transmission efficiency. However, the fidelity of the remotely prepared cat state is always above 0.5 as long as the transmission efficiency in Alice's mode is larger than zero, which is different from that in the case of loss on Bob's channel. This demonstrates that the remotely prepared cat state can tolerate much more loss in Alice's channel than that in Bob's channel.

In our experiment, only the loss in Alice's mode affects the single-photon detection rate. In the case of loss in Bob's channel, the generation rate of optical cat states is around $1$ kHz, which is obtained by considering the single-photon detection rate $14$ kHz and the success probability $7.5\%$ of the post-selection procedure, and remains unchanged with the transmission efficiency $\eta_{B}$.  In the case of loss on Alice's channel, the generation rate of optical cat states decreases with transmission efficiency $\eta_{A}$. The success probability of the post-selection procedure depends on the selection width, which is chosen as $\delta x=0.05$ in our experiment. To choose a proper selection width in post-selection, the trade-off between the fidelity and the success probability should be considered (see Appendix C for more details).

\section{Discussion and Conclusion}

Up to now, it remains a challenge to prepare optical cat states with amplitude larger than $2$, which is a necessary requirement for quantum computation with cat states~\cite{RalphPRA2003}. Here, we show that our scheme can generate large-amplitude optical cat states by subtracting more photons from a TMSS with optimum squeezing. Supposing that $n$ photons are subtracted and the homodyne projective measurement $p_A = 0$ is performed at Alice's station, Bob's state is expressed by 
\begin{equation}
	|\varphi^{(n)}_{0|\hat{p}_{A}}\rangle_{B}=N_n^{-1/2} \sum_{m=n}^{\infty} \frac{\sqrt{m!} i^{-(m-n)}\tanh^m r}{\sqrt{2^{m-n}}(m-n)!}H_{m-n}(0)|m\rangle_{B},
\end{equation}
where $N_n$ is the normalized parameter. When even and odd numbers of photons are subtracted, even and odd cat states are obtained, respectively. The fidelity between Bob's state and an ideal cat state is given by
\begin{equation}
	F^{(n)}_{\pm}=
	\mathcal{N}'_n\left| \sum_{m=n}^{\infty} \frac{ i^{-(m-n)} \tanh^m r}{\sqrt{2^{m-n}}} \frac{\alpha^m \pm (-\alpha)^m}{(m-n)!}H_{m-n}(0)\right|^2,
\end{equation}
where $\mathcal{N}'_n=N_n^{-1}{e^{-|\alpha|^2}}/{2(1\pm e^{-2|\alpha|^2})}$, and the subscripts $+ $ and $-$ correspond to even and odd cat states, respectively.

As shown in Figure~\ref{fig5}a, the amplitude of the cat state is increasing with the increase of squeezing level while the fidelity decreases slowly when one photon is subtracted by Alice (solid lines). Interestingly, by subtracting more photons, the amplitudes are increased but the fidelities reach their maximum at certain squeezing levels, which is different from the tendency of the case of single-photon subtraction. As shown in Figure~\ref{fig5}b, it is obvious that with optimum squeezing level, the fidelity reaches the maximum for each case, the more photons are subtracted from Alice's state, the larger cat states are obtained. For example, when the squeezing level of the TMSS is $8.4$ dB, by subtracting three photons, the cat state with amplitude of $\sim2.61$ and fidelity of $96\%$ can be obtained. By comparing the Wigner functions of subtracting two, three, and four photons, as shown in Figure~\ref{fig5}c, \ref{fig5}d, and \ref{fig5}e respectively, the amplitude of the cat states at Bob's station is increased and the interference between two coherent components becomes more apparent. Compared with the preparation of a large-amplitude cat state by subtracting multi-photons from a squeezed vacuum state, cat states prepared by our method present higher fidelity under the same condition (see Appendix A for more details).

Compared with the method to prepare non-Gaussian entangled states such as N00N state and hybrid entangled state, Gaussian entangled states can be prepared deterministically and are scalable. The scalability of the Gaussian entangled state has been demonstrated with over ten thousand modes in recent years \cite{Mikkel2019,Asavanant2019}. In a quantum network, with the increase of the number of users, it is convenient to extend our RSP scheme to multi-users based on a deterministic multipartite Gaussian entangled state \cite{Han2022}. It is interesting to demonstrate the RSP of optical cat states based on multipartite Gaussian entanglement, which has potential application in a quantum network and is worthy of further investigation.

In summary, we remotely prepared odd optical cat states by subtracting one photon from one mode of the TMSS and performing homodyne projective measurement on the photon-subtracted state at Alice's station. The rotation operation is also implemented on the prepared cat states remotely by changing Alice’s measurement basis of homodyne detection. We demonstrate that the remotely prepared cat state can tolerate much more loss in Alice's channel than that in Bob's channel. More importantly, we show that optical cat states with amplitudes larger than 2 can be prepared by subtracting more photons from the TMSS with optimum squeezing. 

In our scheme, the techniques of photon subtraction and homodyne projective measurement are combined to realize remote preparation of optical cat state, which is a typical hybrid quantum information processing involving both discrete-variable technique and continuous-variable quantum resource. Our results present a new method to remotely prepare optical cat states and make a crucial step toward the remote hybrid quantum information processing. Inspired by the recent advance of preparing non-Gaussian quantum states by performing photon subtraction on a multimode Gaussian entangled state~\cite{Nicolas2019}, it would be worth to further developing our method for remotely creating cat states in a quantum network.

\section{ACKNOWLEDGMENTS}

D. H. and F. S. contributed equally to this work. This research was supported by the NSFC (Grant Nos. 11834010, 12125402, 11975026 and 12147148). X. S. acknowledges the Fund for Shanxi ``1331 Project" Key Subjects Construction. Q. H. acknowledges the Beijing Natural Science Foundation (Z190005). F. S. acknowledges the China Postdoctoral Science Foundation (Grant No. 2020M680186). 

\section*{APPENDIX A: THEORETICAL ANALYSIS OF THE REMOTE STATE PREPARATION PROPOSAL}

\textbf{\newline{1. The case of single-photon subtraction}}\\

Remote state preparation (RSP) of optical cat states is based on the remotely distributed quantum entanglement between Alice and Bob. A two-mode squeezed vacuum state (TMSS) is expressed as $|\psi_0\rangle_{AB}=\frac{1}{\cosh r}\sum^{\infty}_{m=0}\tanh^{m}r|m,m\rangle_{AB}$, where $r$ is the squeezing parameter and $|m\rangle$ is the Fock state. After Alice performs single-photon subtraction operation on mode A, the shared state can be rewritten as
\begin{align}
	|\psi_1\rangle_{AB}&\propto \hat{a}_A|\psi_0\rangle_{AB}= N_1^{-1/2} \sum_{m=1}^{\infty}\sqrt{m}\tanh^{m}r|m-1,m\rangle_{AB}.
	\label{SPS}
\end{align}
where $\hat{a}_A$ is the annihilation operator on Alice's state, and $N_1=\sinh^2r\cosh^2r$ is the normalized parameter. It has been shown that the Bob's state becomes non-Gaussian after Alice's photon subtraction operation~\cite{walschaers2020remote}.

\begin{figure}[h]
\renewcommand*{\thefigure}{A\arabic{figure}}
\includegraphics[width=0.5\textwidth]{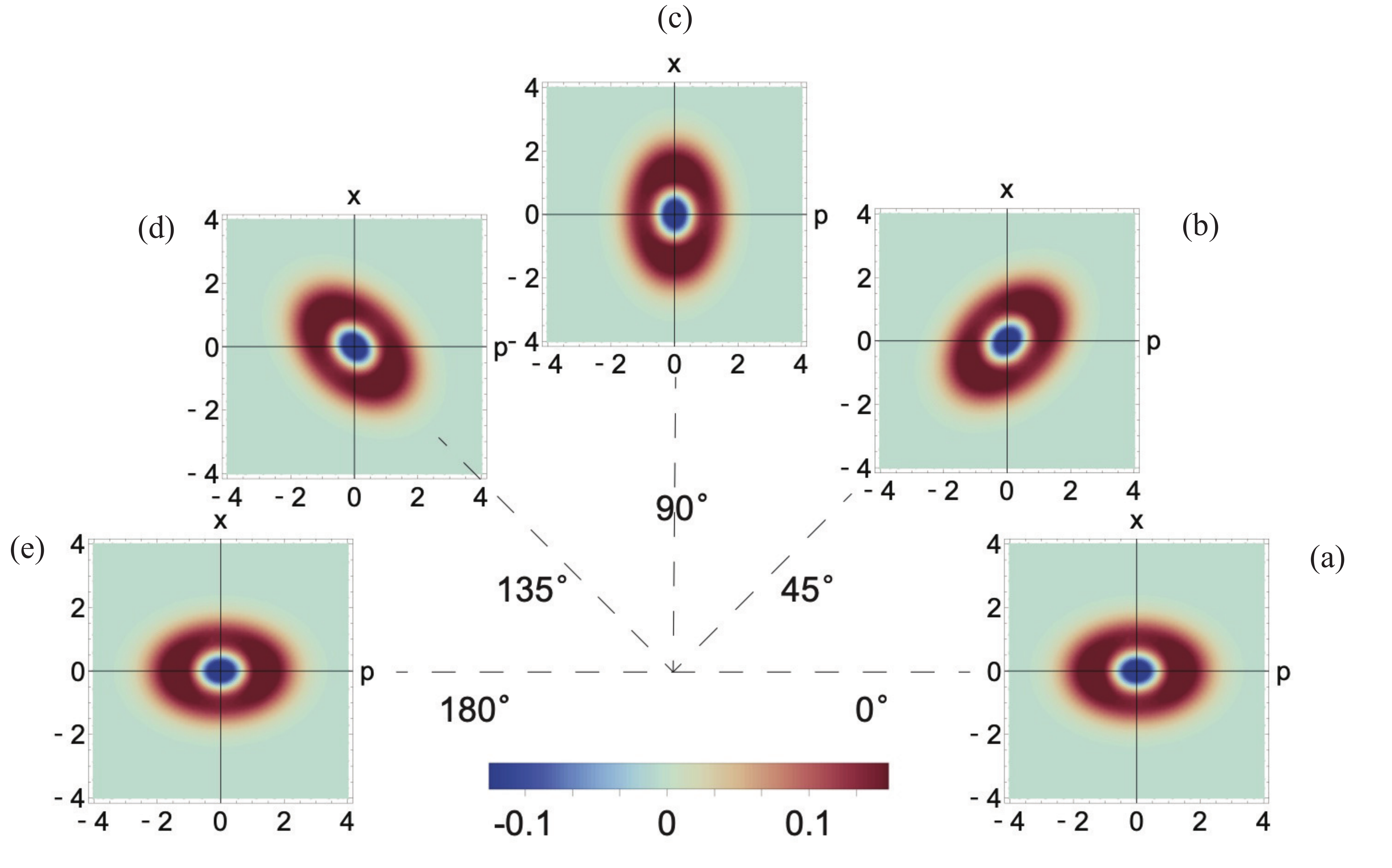}
\caption{Rotated cat states in arbitrary directions. (a-e) Theoretically plot of remote prepared cat states rotated in five directions based on a TMSS with unit transmission efficiency.}
\label{A1}
\end{figure}

By performing the quadrature measurement $\hat{x}_{A}^{\theta}=(\hat{a}_A e^{-i\theta}+\hat{a}_A^{\dagger}e^{i\theta})/\sqrt{2}$ on the photon-subtracted state at Alice's station,
the conditional state at Bob's station becomes $|\varphi^{(1)}_{x^{\theta}_{A}|\hat{x}^{\theta}_{A}}\rangle_{B}={}_{A}\langle x_{A}^{\theta}|\psi_1 \rangle_{AB}$, where $x_A^\theta$ is the outcome of Alice's quadrature measurement. The inner products of the eigenstates of operator $\hat{x}_{A}^{\theta}$ and the photon number states $|m\rangle$ are given by
\begin{equation}
	{_{A}}\langle x_{A}^{\theta}|m \rangle_{A}=\frac{e^{-im\theta}}{\sqrt{2^{m}m!\sqrt{\pi}}}e^{-(x_{A}^{\theta})^2/2}H_m(x_{A}^{\theta}),
	\label{eigenstate}
\end{equation}
where $H_m(x)$ is the Hermite polynomial. Particularly, $\hat{x}_{A}^{\theta}=\hat{x}_{A}$ when $\theta=0$ and $\hat{x}_{A}^{\theta}=\hat{p}_{A}$ when $\theta=\pi/2$, which correspond to the amplitude and phase quadrature operators, respectively. Their corresponding eigenstates lead to
\begin{align}
{_{A}}\langle x_A|m \rangle_{A}&=\frac{1}{\sqrt{2^{m}m!\sqrt{\pi}}}e^{-x_A^2/2}H_m(x_A),
\\ {_{A}}\langle p_A|m \rangle_{A}&=\frac{(-i)^m}{\sqrt{2^{m}m!\sqrt{\pi}}}e^{-p_A^2/2}H_m(p_A).
\end{align}
Therefore, Bob's conditional state becomes an odd cat state with the outcome $x_{A}^{\theta}=0$, which takes the form of
\begin{align}
	 |\varphi^{(1)}_{0|\hat{x}^{\theta}_{A}}\rangle_{B}&=N_B^{-1/2} \sum_{m=1}^{\infty} \frac{\sqrt{m} e^{-i(m-1)\theta}\tanh^m r}{\sqrt{2^{m-1}(m-1)!}}H_{m-1}(0)|m\rangle_{B},
	\label{cat-scheme}
\end{align}
where $N_B$ is the normalized parameter so that ${}_{B}\langle \varphi^{(1)}_{0|\hat{x}_{A}^{\theta}}|\varphi^{(1)}_{0|\hat{x}_{A}^{\theta}} \rangle_{B}=1$. 
Furthermore, considering $H_{2k+1}(0)=0$ and $H_{2k}(0)=(-2)^{k}(2k-1)!!$, Bob's state can be rewritten as
\begin{align}
	 |\varphi^{(1)}_{0|\hat{x}^{\theta}_{A}}\rangle_{B}&\propto|r,\theta\rangle_B^{(1)}-|-r,\theta\rangle_B^{(1)},
\end{align}
where $|r,\theta\rangle_B^{(1)}\propto\sum_{m=0}^{\infty}\frac{m!!}{\sqrt{m!}}\text{tanh}^{m}r e^{-i m(\theta-\pi/2)}|m\rangle_B$. Hence, if Alice measures the phase ($\hat{p}_A$) quadrature, Bob's state is similar with an cat state ($|\alpha\rangle-|-\alpha\rangle$) with $|r,\pi/2\rangle_B^{(1)}\propto\sum_{m=0}^{\infty}\frac{m!!}{\sqrt{m!}}\text{tanh}^{m}r |m\rangle_B$. And if Alice measures the amplitude ($\hat{x}_A$) quadrature, Bob's state is similar with the rotated cat state for 90 degree ($|i\alpha\rangle-|-i\alpha\rangle$) with $|r,0\rangle_B^{(1)}\propto\sum_{m=0}^{\infty}(i)^{m}\frac{m!!}{\sqrt{m!}}\text{tanh}^{m}r |m\rangle_B$. By measuring arbitrary quadrature  $x_{A}^{\theta}$, the prepared optical cat state can be rotated for arbitrary degree at Bob's station remotely, as shown in Fig.~\ref{A1}.

\begin{figure}[t]
\renewcommand*{\thefigure}{A\arabic{figure}}
\includegraphics[width=0.5\textwidth]{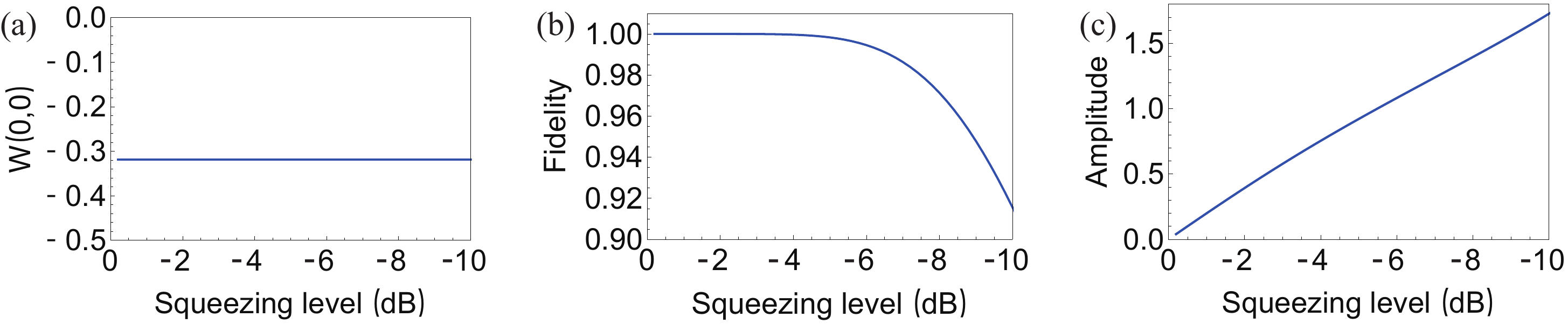}
\caption{Influence of squeezing of TMSS on the prepared cat state in the case of one-photon subtraction. (a) The value of W(0,0). (b) The fidelity. (c) The amplitude of the cat states.}
\label{A2}
\end{figure}

In the following, we examine key figures of merit for the prepared cat state at Bob's station by Wigner function, which represents a typical phase-space quasiprobability distribution and the joint probability distribution of the quadratures for the quantum state~\cite{wigner1932on}. Given the density matrix of the quantum state $\rho$, the corresponding Wigner function is defined as
\begin{equation}
W(X, Y)=\frac{1}{\pi \hbar} \int e^{-2 i x Y / \hbar}\langle X-x|\rho| X+x\rangle d x,
\end{equation} 
with the quadratures expressed by the coherent amplitudes $X=(\alpha+\alpha^{*})/{\sqrt{2}}$, $Y=(\alpha-\alpha^{*})/{\sqrt{2} i}$. If the density matrix is expanded in the Fock basis, $\rho=\sum_{m,n}\rho_{nm}|n\rangle\langle m|$, the Wigner function can be rewritten as~\cite{pathak2014wigner}
\begin{equation}
W\left(\alpha, \alpha^{*}\right)=\sum_{n}^{N_{c}} \rho_{n n} X_{n n}+2 \operatorname{Re}\left(\sum_{m=1}^{N_{c}} \sum_{n=0}^{m-1} \rho_{n m} X_{n m}\right).
\end{equation}
Here, $N_c$ is the large cutoff of photon number, $X_{n m}=\frac{2(-1)^{n}}{\pi} \sqrt{\frac{n !}{m !}} e^{-2|\alpha|^{2}}(2 \alpha)^{m-n} L_{n}^{m-n}\left(4|\alpha|^{2}\right)$, and $L_{b}^{a}(x)$ is the associated Laguerre polynomial.

For an odd cat state, the value of the Wigner function at original point $W(0,0)$ is negative. In Fig.~\ref{A2}(a), the value of $W(0,0)$ is shown with the change of the squeezing level. Since $W(0,0)$ is negative in a large range of the squeezing level, it is verified that a nonclassical state has been prepared on Bob's side, which, as explained below, has a high fidelity with an odd cat state.

An ideal odd cat state is defined as $|cat_{-}\rangle=(\vert\alpha\rangle-\vert-\alpha\rangle)/\sqrt{2(1-e^{-2\vert\alpha\vert^2}})$, where $|\alpha\rangle$ is the coherent state with amplitude $\alpha$.
Since the target state is a pure state, the fidelity between Bob's state $\rho_{B}$ and ideal cat state can be simplified as $F^{(1)}_{-}=\langle cat_{-}|\rho_{B}|cat_{-} \rangle$~\cite{jozsa1994fidelity}.
 The dependence of fidelity and amplitude of Bob's cat state on squeezing level are shown in Fig.~\ref{A2}(b) and \ref{A2}(c), respectively. It is obvious that the cat state with high fidelity can be prepared at Bob's station within a large range of the squeezing level (0 - 6 dB) of the TMSS. However, with the increase of squeezing level, the amplitude is increased while the fidelity drops slightly.
 
\textbf{\newline{2. The case of two-photon subtraction}}\\

The expression of Bob's state when $n$ photons are subtracted from Alice's mode are presented in the main text, which illustrate that odd or even cat states can be prepared remotely by subtracting odd or even photon numbers at Alice's station. To show the feasibility of preparing an even cat state in our scheme, the example of $n = 2$ is presented. The TMSS after Alice performs two-photon subtraction on her mode becomes $|\psi_2\rangle_{AB}\propto \hat{a}_A^2|\psi_0\rangle_{AB}\propto \sum_{m=2}^{\infty}\sqrt{m(m-1)}\tanh^{m}r|m-2,m\rangle_{AB}$. After the homodyne projective measurement performed by Alice, Bob's state can be rewritten as
\begin{align}
	 |\varphi^{(2)}_{0|\hat{x}^{\theta}_{A}}\rangle_{B}&\propto \sum_{m=2}^{\infty} \frac{\sqrt{m(m-1)} e^{-i(m-2)\theta}\tanh^m r}{\sqrt{2^{m-2}(m-2)!}}H_{m-2}(0)|m\rangle_{B}\nonumber\\
	&\propto |r,\theta\rangle^{(2)}_B+|-r,\theta\rangle^{(2)}_B,
	\label{eq:cat-even}
\end{align}
where $|r,\theta\rangle^{(2)}_B\propto\sum_{m=1}^{\infty}\frac{m(m-1)!!}{\sqrt{m!}}\text{tanh}^{m}r e^{-i m(\theta-\pi/2)}|m\rangle_B$. It's obvious that Eq.~(\ref{eq:cat-even}) takes a similar form with the even cat state $|cat_{+}\rangle=(|\alpha\rangle+|-\alpha\rangle)/\sqrt{2(1+e^{-2|\alpha|^2})}$. As explained in the case of the odd cat state, an even cat state can be prepared at Bob's station if Alice take the phase ($\hat{p}_A$) quadrature measurement with $|r,\pi/2\rangle^{(2)}_B\propto\sum_{m=1}^{\infty}\frac{m(m-1)!!}{\sqrt{m!}}\text{tanh}^{m}r |m\rangle_B$, and an rotated cat state for 90 degree can be generated at Bob's station if Alice take the amplitude ($\hat{x}_A$) quadrature measurement with $|r,0\rangle^{(2)}_B\propto\sum_{m=1}^{\infty}(i)^m\frac{m(m-1)!!}{\sqrt{m!}}\text{tanh}^{m}r |m\rangle_B$.

\textbf{\newline{3. Comparison of two schemes based on a SMSS and a TMSS}}\\

In the last decades, local preparation of cat state by subtracting photons from a single-mode squeezed state (SMSS) has been well developed~\cite{Ourjoumtsev2006,Neergaard2006,kentaro2007,Takahash2008,Thomas2010,Sychev2017,zhang2021}. To show the advantage of our scheme in the preparation of large-amplitude cat state, we compare the scheme of preparing cat states based on a TMSS with this commonly used method, i.e., subtracting photons from a SMSS.

\begin{figure}[t]
\renewcommand*{\thefigure}{A\arabic{figure}}
\includegraphics[width=0.47\textwidth]{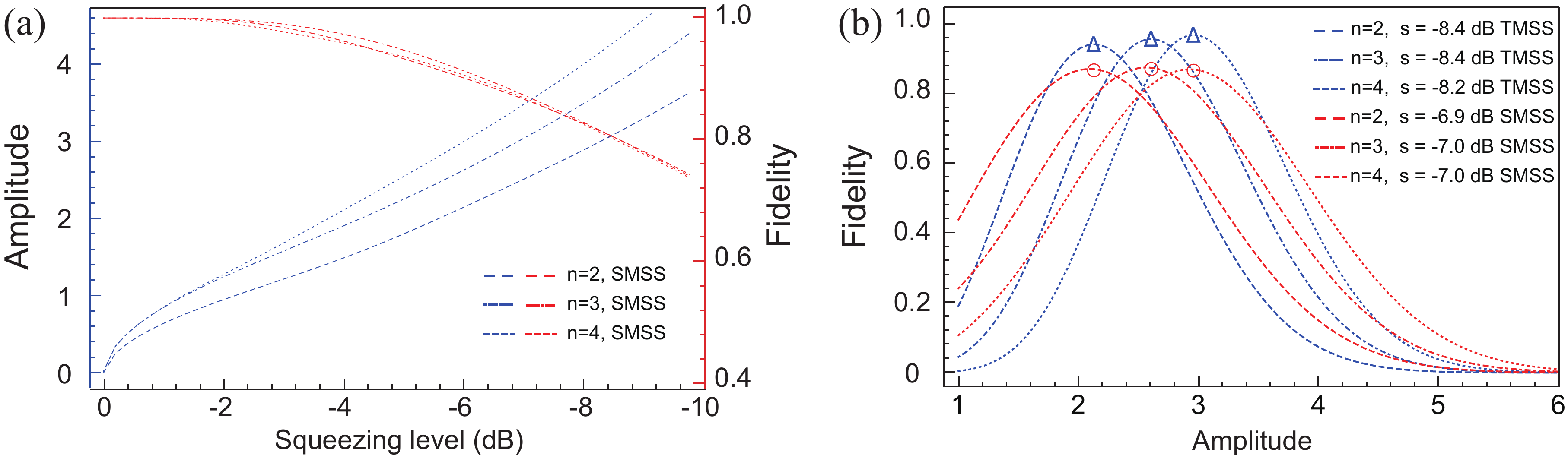}
\caption{Comparation of the results in two schemes based on a SMSS and a TMSS. (a) Dependence of amplitude (blue lines) and fidelity (red lines) of the locally prepared cat state on the squeezing level for different subtracted photon numbers in the scheme of subtracting photons from a SMSS. (b)  Dependence of fidelities on the amplitude of the cat state with optimum squeezing level for different subtracted photon numbers of two schemes. The red curves with circles represents the cat states prepared locally by subtracting photons from the SMSS, while the blue curves with triangles represents the cat states by subtracting photons from the TMSS.}
\label{A3}
\end{figure}

A SMSS in the Fock basis is expressed by,
\begin{equation}
|\phi_0\rangle_B=\frac{1}{\sqrt{\cosh r}}\sum_{m=0}^{\infty}\tanh^m r\frac{\sqrt{(2m)!}}{2^m m!}|2m\rangle_B.
\end{equation}
By subtracting $n$ photons from the SMSS, the state becomes,
\begin{equation}
|\phi_n\rangle_B=\mathcal{N}_{n}^{-1/2}\sum_{m=\lceil \frac{n}{2} \rceil}^{\infty}\tanh^m r\frac{(2m)!}{2^m m!\sqrt{(2m-n)!}}|2m-n\rangle_B.
\end{equation}
Here $\mathcal{N}_{n}$ is the normalized parameter, and the ceiling function $\lceil \frac{n}{2} \rceil$ represents the least integer greater than or equal to $n/2$. The ideal cat states in Fock basis can be expressed by:
\begin{align}
	 |\psi^{\pm}_{\text{cat}}\rangle&=\frac{e^{-|\alpha|^2/2}}{\sqrt{2(1\pm e^{-2|\alpha|^2})}} \sum_{m=0}^{\infty} \frac{\alpha^m \pm (-\alpha)^m}{\sqrt{m!}}|m\rangle,
	 \label{eq:cat}
\end{align}
Therefore, the corresponding fidelity with the ideal cat states (\ref{eq:cat}) can be expressed as,
\begin{equation}
\begin{aligned}
\mathcal{F}_{\pm}^{(n)}=&\mathcal{N}_{n}^{-1} \left|\frac{e^{-|\alpha|^2/2}}{\sqrt{2(1\pm e^{-2|\alpha|^2})}}\right|^2
\\&\times \left|\sum_{m=\lceil \frac{n}{2} \rceil}^{\infty} \frac{ (2m)! \tanh^m r}{2^{m}m!} \frac{\alpha^{2m-n} \pm (-\alpha)^{2m-n}}{(2m-n)!}\right|^2.
\end{aligned}
\end{equation}

As shown in Fig.~\ref{A3}, although the amplitudes of the locally prepared cat states always increase, the fidelities $\mathcal{F}_{\pm}^{(n)}$ always decease, along with the increase of the squeezing level for different subtracted photon numbers. This behavior is clearly different from our scheme, i.e., subtracting photons from a TMSS, when $n\geq 2$, where the fidelities reach the maximum with optimal squeezing levels as shown in Fig.~4 of the main text. We also compare the fidelities of these two schemes when preparing large-amplitude cat states with the same amplitudes, as shown in Fig.~\ref{A3}b. It is obvious that the fidelities of our scheme are always higher than that of subtracting photons from a SMSS in the case of subtracting two, three and four photons. This confirms that our scheme is more suitable for preparing large-amplitude cat states when multi-photons are subtracted. 

\section*{APPENDIX B: DETAILS OF THE EXPERIMENT}

\textbf{\newline{1. Experimental setup}}\\

\begin{figure}[h]
\renewcommand*{\thefigure}{B\arabic{figure}}
\includegraphics[width=0.48\textwidth]{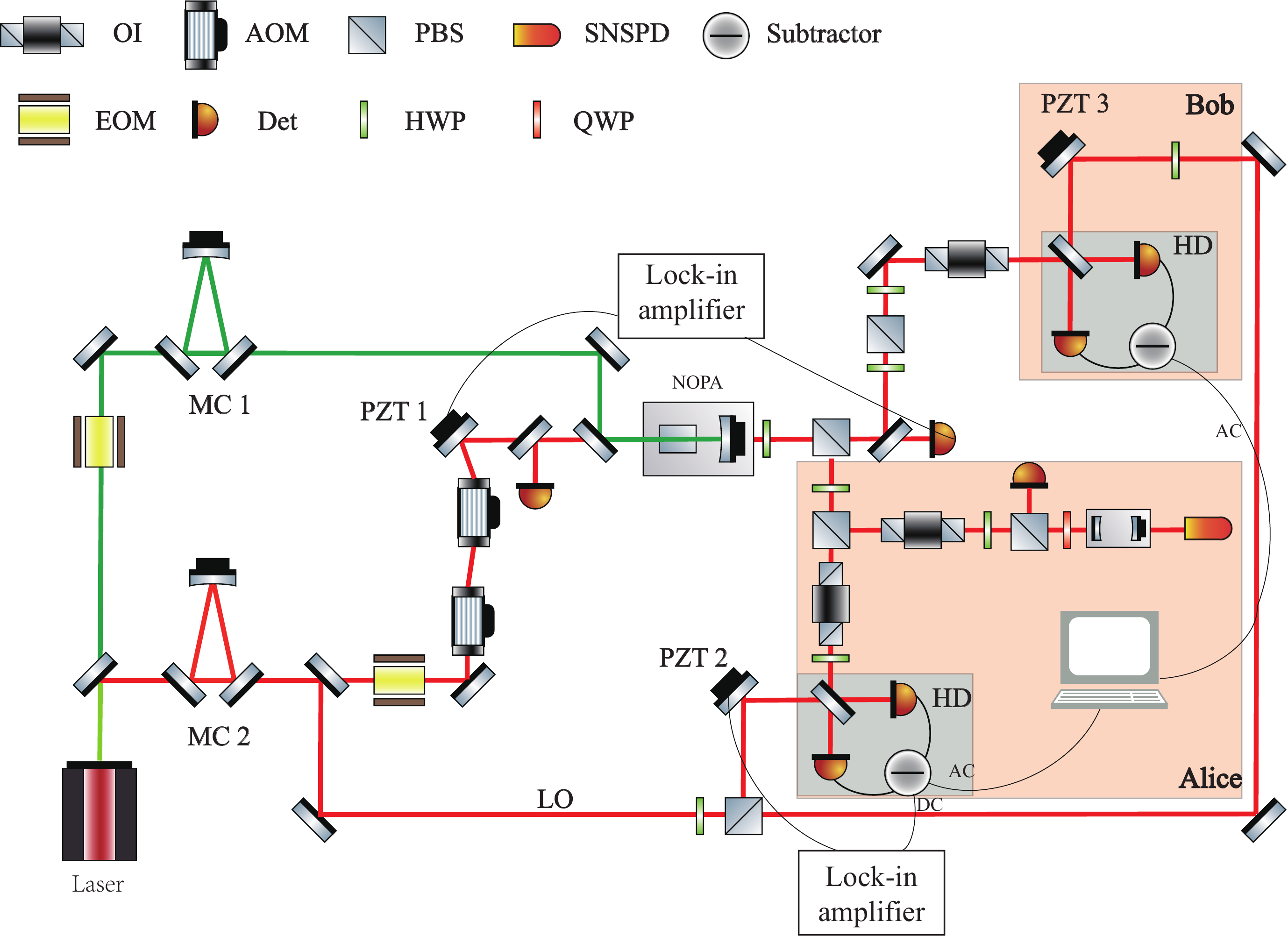}
\caption{Detailed experimental setup. MC: mode cleaner, NOPA: non-degenerate optical parametric amplifier, OI: optical isolator, EOM: electro-optic modulator, AOM: acousto-optic modulator, Det: detector, PBS: polarization beam splitter, HWP: half-wave plate, SNSPD: superconducting nanowire single-photon detector, QWP: quarter-wave plate, HD: homodyne detector, PZT: piezo transducer, LO: local oscillator.}
\label{B1}
\end{figure}

Figure~\ref{B1} shows the detailed experimental setup. A continuous wave intracavity frequency-doubled and frequency-stabilized Nd:YAP/LBO (Nd-doped YAIO3 perorskite lithium triborate) laser generates laser beams at 1080 nm and 540 nm simultaneously, which serve as the seed and pump beams of the NOPA respectively. The NOPA is composed of a 10 mm long $\alpha$-cut type-II potassium titanyl phosphate (KTP) crystal and a concave mirror with a 50 mm radius. The front face of the KTP crystal is coated to be used for the input coupler and the concave mirror serves as the output coupler of NOPA. The transmittances of the front face of KTP crystal at 540 nm and 1080 nm are $40\%$ and $0.04\%$, respectively. The end-face of KTP is antireflection coated for both 1080 nm and 540 nm. The transmittances of the output coupler at 540 nm and 1080 nm are $0.5\%$ and $12.5\%$, respectively. 

In our experiment, to produce the two-mode squeezed state (TMSS), the NOPA is locked with the lock-and-hold technique, which is performed with the help of two acousto-optic modulators (AOMs). The seed beam is chopped into a cyclic form with $50$ ms period, which corresponds to each locking and hold period, using two AOMs. During the locking period when AOMs are switched on, the first order of the AOM transmissions is injected into the NOPA for the cavity locking. When the AOMs are switched off, the seed beam is chopped off and the NOPA is holding. The TMSS is generated and the measurement is performed during the hold period. The filter cavity used to filter out the non-degenerate mode is also locked during the lock period by using the signal beam transmitted through the NOPA. The reflected light of the input coupler of the filter cavity is detected and the error signal is obtained by using a lock-in amplifier. The relative phase of Alice's HD is locked to 0 degrees and 90 degrees to measure amplitude and phase quadratures, respectively. The relative phase of Bob's HD is scanned to reconstruct the Wigner function of Bob's mode by using the maximum-likelihood method.

The total detection efficiency of HD is around $90\%$ which includes the quantum efficiency of the photodiode ($98\%$), the mode matching efficiency ($98\%$), and the clearance of the HD ($96\%$). The transmission efficiencies of two optical isolators together with other optical elements placed in front of Alice's and Bob's HDs lead to $\sim10\%$ transmission loss. Thus the cat state with the maximum transmission efficiency of 0.9 is obtained in the experiment. 

\textbf{\newline{2. Covariance matrix of the TMSS}}\\

The initial TMSS is characterized by reconstructing its covariance matrix (CM). The theoretical CM without loss is expressed by
\begin{equation}
\sigma_{A B}
=\left(\begin{array}{cccc}
\Delta^{2}\hat{x}_{A}& 0 & \Delta^{2}(\hat{x}_A\hat{x}_B) & 0 \\ 
0 & \Delta^{2}\hat{p}_{A} & 0 & \Delta^{2}(\hat{p}_A\hat{p}_B)  \\ 
\Delta^{2}(\hat{x}_A\hat{x}_B)  & 0 & \Delta^{2}\hat{x}_{B} & 0 \\ 
0 & \Delta^{2}(\hat{p}_A\hat{p}_B)  & 0 & \Delta^{2}\hat{p}_{B} 
\end{array}\right).
\end{equation}
where $\Delta^{2}\hat{x}_{A}$ = $\Delta^{2}\hat{p}_{A}$ = $\Delta^{2}\hat{x}_{B}$ = $\Delta^{2}\hat{p}_{B}$ = $\frac{V_a+V_s}{2}$, and $\Delta^{2}(\hat{x}_A\hat{x}_B)$ = $\frac{V_a-V_s}{2}$, $\Delta^{2}(\hat{p}_A\hat{p}_B)$ = $-\frac{V_a-V_s}{2}$, $V_a=\Delta^{2}(\hat{x}_A-\hat{x}_B)/2=\Delta^{2}(\hat{p}_A+\hat{p}_B)/2$ and $V_s=\Delta^{2}(\hat{x}_A+\hat{x}_B)/2=\Delta^{2}(\hat{p}_A-\hat{p}_B)/2$ are the correlated variances of the quadrature measurement statistics between two modes of the TMSS. In our experiment, $V_{s}=0.24$  and $V_{a}=1.3$, which correspond to $-3.2$ dB squeezing and $4.2$ dB antisqueezing when the NOPA is pumped at $70$ mW, respectively.

In the process of measuring the covariance matrix of the TMSS, the variances of quadrature phase $\hat{p}_{A(B)}$ and amplitude $\hat{x}_{A(B)}$ of the TMSS state are measured by locking the relative phase between the local beams and signal beams of two HDs to 90 and 0 degrees, respectively. A digital storage oscilloscope (OSC, TELEDNE LECROY, HDO8108A) with sampling rate of 500 MS/s is used to recored the electrical signals from two HDs. The corresponding variances $\Delta^{2}\hat{p}_{A(B)}$ and $\Delta^{2}\hat{x}_{A(B)}$ are calculated from the measured data, as well as the correlated variances $\Delta^{2}(\hat{x}_A-\hat{x}_B)$ and $\Delta^{2}(\hat{p}_A+\hat{p}_B)$. Cross correlations are obtained by using the relations $\Delta^{2}(\hat{x}_A\hat{x}_B)=(\Delta^{2}(\hat{x}_A-\hat{x}_B)-\Delta^{2}\hat{x}_A-\Delta^{2}\hat{x}_B)/2$ and  $\Delta^{2}(\hat{p}_A\hat{p}_B)=(\Delta^{2}(\hat{p}_A+\hat{p}_B)-\Delta^{2}\hat{p}_A-\Delta^{2}\hat{p}_B)/2$, respectively.

\section*{APPENDIX C: THE EFFECT OF THE EXPERIMENTAL IMPERFECTIONS AND TRANSMISSION EFFICIENCY}

Taking the experimental imperfections into account, the initial entangled state generated by Alice is actually a mixed two-mode squeezed state, which can be described by the variances of squeezing ($V_s$) and antisqueezing ($V_a$) with $V_sV_a>1/4$. Thus, a mixed two-mode Gaussian entangled state is shared by Alice and Bob in the practical situation, whose CM can be expressed by
\begin{equation}
\sigma_{A B}=
\left(\begin{array}{cc}\sigma_{A} & \gamma_{A B} \\ \gamma_{A B}^{\top} & \sigma_{B}\end{array}\right)
=\left(\begin{array}{cccc}
n & 0 & c_{1} & 0 \\ 
0 & n & 0 & c_{2} \\ 
c_{1} & 0 & m & 0 \\ 
0 & c_{2} & 0 & m 
\end{array}\right),
\label{cm}
\end{equation}
where $n=\Delta^{2}\hat{x}_{A}=\Delta^{2}\hat{p}_{A}$, $m=\Delta^{2}\hat{x}_{B}=\Delta^{2}\hat{p}_{B}$ represent the variances of amplitude and phase quadratures of the output optical modes, $c_{1}=Cov(\hat{x}_{A},\hat{x}_{B})$ and $c_{2}=Cov(\hat{p}_{A},\hat{p}_{B})$ indicate their cross correlations.

Considering the practical losses characterized by transmission efficiency $\eta_A$ and $\eta_B$,  the CM elements become $n=\eta_A(V_a+V_s)/2+(1-\eta_A)/2$, $m=\eta_B(V_a+V_s)/2+(1-\eta_B)/2$, $c=\sqrt{\eta_A\eta_B}(V_a-V_s)/2$. In our experiment, the loss introduced by the isolator at Alice's station also need to be considered, which lead to $\eta_A = 0.9$. The squeezing level is defined as $s=-10\log_{10}(2V_s)$ dB.

In the following, we show that such a two-mode Gaussian entangled state shared by Alice and Bob can be mapped to the state with an effective model in Ref.~\cite{xiang2017investigating}. In the effective model, one mode of an effective TMSS freely propagates towards Bob while the other mode propagates towards Alice through a phase-insensitive loss channel with transmissivity $\eta$ and a two-mode squeezer with a squeezing parameter $r_s$. Thus, the CM of the two-mode state takes the form of~\cite{xiang2017investigating}
\begin{equation}
	\sigma_{\text{mixed}}=\frac{1}{2}\left( 
		\begin{array}{cccc}
			(\tau b+\xi) & 0 & \sqrt{\tau(b^2-1)} & 0 \\
			0 & (\tau b+\xi) & 0 & -\sqrt{\tau(b^2-1)} \\
			\sqrt{\tau(b^2-1)} & 0 & b & 0 \\
			0 & -\sqrt{\tau(b^2-1)} & 0 & b
		\end{array}
	\right),
	\label{CM_TMST}
\end{equation}
with $\tau=\eta \cosh^2(r_s)$, $\xi=(1-\eta)\cosh^2(r_s)+\sinh^2(r_s)$, $b=(1+\zeta^2)/(1-\zeta^2)$, where $\zeta$ is the squeezing parameter of the effective TMSS. The corresponding density matrix can be expressed in the Fock basis as follows~[$40$]
\begin{eqnarray}
\rho_{\text{mixed}}\propto&\sum_{m_1,m_2,n_1,n_2=0}^{\infty}\delta_{m_1+n_2,m_2+n_1}\frac{1-\zeta^2}{\cosh^2r_s}\left( \frac{\zeta\sqrt{\eta}}{\cosh r_s} \right)^{m_2+n_2}\nonumber\\
\times&\tanh^{2(m_1-m_2)}r_s\nonumber\sum_{k=\max\left\{ 0,m_2-m_1 \right\}}^{\min\left\{ m_2,n_2 \right\}}\sqrt{\binom{m_2}{k}\binom{n_2}{k}\binom{m_1}{m_2-k}\binom{n_1}{n_2-k}}\nonumber\\
\times&\left( \sqrt{\frac{1-\eta}{\eta}}\sinh r_s \right)^{2k}|m_1m_2\rangle\langle n_{1}n_{2}|.
\label{TMST}
\end{eqnarray}

\begin{figure}[t]
\renewcommand*{\thefigure}{C\arabic{figure}}
\centering
\includegraphics[width=0.48\textwidth]{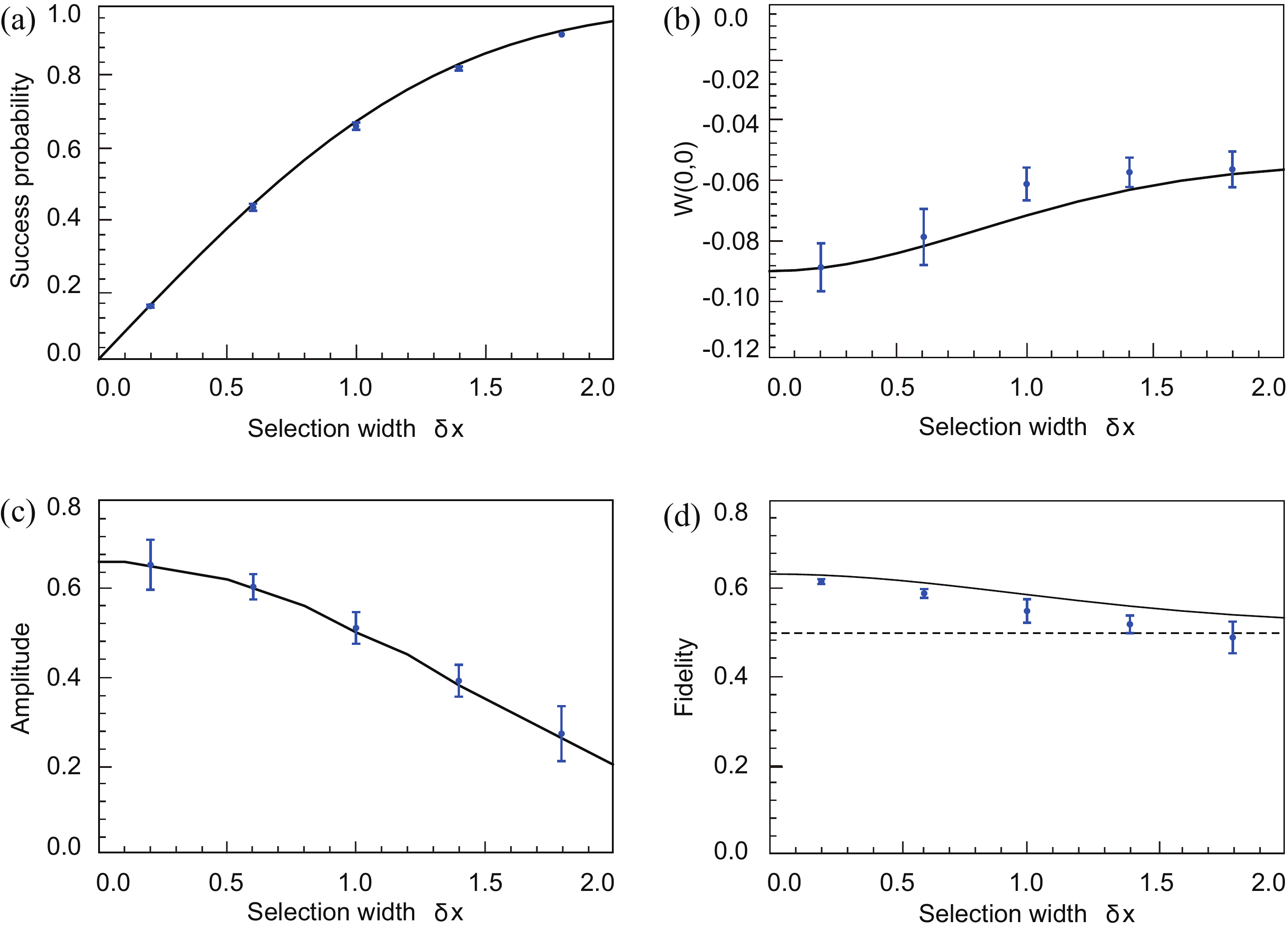}
\caption{The influence of the selection width in the homodyne projective measurement. (a) The successful probability $P$, (b) the value of $W(0,0)$, (c) the amplitude of cat state $|\alpha|$, and (d) the fidelity $F$ changing with the selection width $\delta x$. The error bars are obtained by standard deviation of measurements repeated three times.}
\label{C1}
\end{figure}
It is well known that Gaussian states are fully characterized by the CM of their quadrature observables, when the mean values are all set as zeros.
Comparing the CMs in Eq.~(\ref{cm}) and Eq.~(\ref{CM_TMST}), they are identical if the following conditions are satisfied
\begin{align}
2n&=\tau b+\xi,\;	2m=b,\; 2c=\sqrt{\tau(b^2-1)}.
\label{eq:satisfied}
\end{align}
Hence, the density matrices of the states with identical CMs~(\ref{cm}) and (\ref{CM_TMST}) are actually the same. This means that the two-mode Gaussian entangled state shared by Alice and Bob can be expressed by Eq.~(\ref{TMST}) with the parameters determined by Eqs.~(\ref{eq:satisfied}).

With such a mixed state shared by Alice and Bob, the single-photon subtraction operation on Alice’s state, $\rho_{\text{mixed}}^{sub}=\hat{a}_A \rho_{\text{mixed}} \hat{a}_A^\dagger$, leads to the density matrix of the following form
\begin{eqnarray}
	\rho_{\text{mixed}}^{sub}&\propto\sum_{m_1,n_1=1}^{\infty}\sum_{m_2,n_2=0}^{\infty}\delta_{m_1+n_2,m_2+n_1}\frac{1-\zeta^2}{\cosh^2r_s}\left( \frac{\zeta\sqrt{\eta}}{\cosh r_s} \right)^{m_2+n_2}\nonumber\\
	\times & \sqrt{m_1n_1} \tanh^{2(m_1-m_2)}r_s\nonumber\sum_{k=\max\left\{ 0,m_2-m_1 \right\}}^{\min\left\{ m_2,n_2 \right\}} \sqrt{\binom{m_2}{k}\binom{n_2}{k}\binom{m_1}{m_2-k}\binom{n_1}{n_2-k}}\\
	\times &\left( \sqrt{\frac{1-\eta}{\eta}}\sinh r_s \right)^{2k} |m_1-1,m_2\rangle\langle n_{1}-1,n_{2}|.
	\label{TMST-sub}
\end{eqnarray}
Then by performing projective measurement $\hat{x}^{\theta}_{A}$ on Alice’s state and obtaining the outcome of $x_A^\theta$, the Bob's state $\rho_{\text{mixed}}^{B}=\langle x_A^\theta|\rho_{\text{mixed}}^{sub}|x_A^\theta \rangle$ collapses to
\begin{eqnarray}
	\rho_{\text{mixed}}^{B}&\propto\sum_{m_1,n_1=1}^{\infty}\sum_{m_2,n_2=0}^{\infty}\delta_{m_1+n_2,m_2+n_1}\frac{1-\zeta^2}{\cosh^2r_s}\left( \frac{\zeta\sqrt{\eta}}{\cosh r_s} \right)^{m_2+n_2}\nonumber\\
	 \times&\frac{e^{i(n_1-m_1)\theta}e^{-(x_A^\theta)^2}\sqrt{m_1n_1}}{\sqrt{2^{m_1+n_1-2}(m_1-1)!(n_1-1)!\pi}}H_{m_1-1}(x_A^\theta) H_{n_1-1}(x_A^\theta)\nonumber\\
	  \times&\tanh^{2(m_1-m_2)}r_s\sum_{k=\max\left\{ 0,m_2-m_1 \right\}}^{\min\left\{ m_2,n_2 \right\}}\sqrt{\binom{m_2}{k}\binom{n_2}{k}\binom{m_1}{m_2-k}\binom{n_1}{n_2-k}}\nonumber\\
	\times& \left( \sqrt{\frac{1-\eta}{\eta}}\sinh r_s \right)^{2k}|m_2\rangle\langle n_{2}|.
	\label{TMST-cat}
\end{eqnarray}

The theoretical curves of the value of $W(0,0)$, fidelity and amplitude in Fig.~3 in the main text are all based on the Bob's state shown in Eq. (\ref{TMST-cat}).

In our experiment, we have assumed the outcomes in a narrow range $x_A^\theta\in[-\delta x,\delta x]$ can be regarded as $x_A^\theta=0$, where a finite successful probability $\sim7.5\%$ with $\delta x=0.05$ is obtained in the main text.
Specifically, the success probability density to obtain the outcome of $x_A^\theta$ is expressed as
\begin{eqnarray}
p(x_A^\theta)&=M^{-1}\sum_{n_1=1}^{\infty}\sum_{n_2=0}^{\infty}\frac{1-\zeta^2}{\cosh^2r_s}\left( \frac{\zeta\sqrt{\eta}}{\cosh r_s} \right)^{2n_2} \frac{e^{-(x_A^\theta)^2}n_1}{2^{n_1-1}(n_1-1)!\sqrt{\pi}} \nonumber\\ 
\times&H^2_{n_1-1}(x_A^\theta) \tanh^{2(n_1-n_2)}r_s \nonumber\\ \times&\sum_{k=\max\left\{ 0,n_2-n_1 \right\}}^{n_2}\binom{n_2}{k}\binom{n_1}{n_2-k}\left( \sqrt{\frac{1-\eta}{\eta}}\sinh r_s \right)^{2k},
\label{pd}
\end{eqnarray}
where $M$ is a normalized parameter to make $\int_{-\infty}^{\infty}p(x_A^\theta)dx_A^\theta=1$. Thus the success probability to obtain outcomes in $x_A^\theta\in[-\delta x,\delta x]$ on mode A is $P(\delta x)=\int_{-\delta x}^{\delta x}p(x_A^\theta)dx_A^\theta$. In the meanwhile, the corresponding density matrix takes the form of

\begin{eqnarray}
\rho_{\text{mixed}}^{B,\delta x}&\propto&\int_{-\delta x}^{\delta x}dx_A^\theta\;\sum_{m_1,n_1=1}^{\infty}\sum_{m_2,n_2=0}^{\infty}\delta_{m_1+n_2,m_2+n_1}\frac{1-\zeta^2}{\cosh^2r_s}\left( \frac{\zeta\sqrt{\eta}}{\cosh r_s} \right)^{m_2+n_2} \nonumber
\\ &\times& \frac{e^{i(n_1-m_1)\theta}\sqrt{m_1n_1}}{\sqrt{2^{m_1+n_1-2}(m_1-1)!(n_1-1)!\pi}}e^{-(x_A^\theta)^2}H_{m_1-1}(x_A^\theta)H_{n_1-1}(x_A^\theta)\nonumber
\\ &\times& \tanh^{2(m_1-m_2)}r_s\nonumber \sum_{k=\max\left\{ 0,m_2-m_1 \right\}}^{\min\left\{ m_2,n_2 \right\}}\sqrt{\binom{m_2}{k}\binom{n_2}{k}\binom{m_1}{m_2-k}\binom{n_1}{n_2-k}}
\\ &\times &\left( \sqrt{\frac{1-\eta}{\eta}}\sinh r_s \right)^{2k}|m_2\rangle\langle n_{2}|.
\label{TMST-cat-delta}
\end{eqnarray}

In Fig.~\ref{C1}, the analytical results of success probability, the value of $W(0,0)$, amplitude and fidelity changing with the selection width $\delta x$ are displayed based on Eqs.~(\ref{pd}) and (\ref{TMST-cat-delta}), which are compared with the experimental data shown as the dots with error bars.
By extending the selection width $\delta x$, the success probability increases while the both of the amplitude and the fidelity decrease. Therefore, Alice needs to choose a proper selection width $\delta x$ in order to remotely prepare a cat state with high fidelity and high success probability at Bob's side.

\end{document}